\documentstyle[preprint,tighten,epsf,aps]{revtex}
\begin{document}
\draft
\title{
Search for High Mass Top Quark Production in $p\bar p$ Collisions
at $\sqrt{s} = 1.8$ TeV
}

%
\author{
S.~Abachi,$^{12}$
B.~Abbott,$^{32}$
M.~Abolins,$^{22}$
B.S.~Acharya,$^{38}$
I.~Adam,$^{10}$
D.L.~Adams,$^{33}$
M.~Adams,$^{15}$
S.~Ahn,$^{12}$
H.~Aihara,$^{19}$
G.~\'{A}lvarez,$^{16}$
% ...
\\
\vskip 0.25cm
\centerline{(D\O\ Collaboration)}
\vskip 0.25cm
}
\address{
\centerline{$^{1}$Universidad de los Andes, Bogota, Colombia}
\centerline{$^{2}$University of Arizona, Tucson, Arizona 85721}
\centerline{$^{3}$Brookhaven National Laboratory, Upton, New York 11973}
\centerline{$^{4}$Brown University, Providence, Rhode Island 02912}
\centerline{$^{5}$University of California, Davis, California 95616}
\centerline{$^{6}$University of California, Irvine, California 92717}
\centerline{$^{7}$University of California, Riverside, California 92521}
\centerline{$^{8}$LAFEX, Centro Brasileiro de Pesquisas F{\'\i}sicas,
                  Rio de Janeiro, Brazil}
\centerline{$^{9}$CINVESTAV, Mexico City, Mexico}
\centerline{$^{10}$Columbia University, New York, New York 10027}
\centerline{$^{11}$Delhi University, Delhi, India 110007}
\centerline{$^{12}$Fermi National Accelerator Laboratory, Batavia,
                   Illinois 60510}
\centerline{$^{13}$Florida State University, Tallahassee, Florida 32306}
\centerline{$^{14}$University of Hawaii, Honolulu, Hawaii 96822}
\centerline{$^{15}$University of Illinois, Chicago, Illinois 60680}
\centerline{$^{16}$Indiana University, Bloomington, Indiana 47405}
\centerline{$^{17}$Iowa State University, Ames, Iowa 50011}
\centerline{$^{18}$Korea University, Seoul, Korea}
\centerline{$^{19}$Lawrence Berkeley Laboratory, Berkeley, California 94720}
\centerline{$^{20}$University of Maryland, College Park, Maryland 20742}
\centerline{$^{21}$University of Michigan, Ann Arbor, Michigan 48109}
\centerline{$^{22}$Michigan State University, East Lansing, Michigan 48824}
\centerline{$^{23}$Moscow State University, Moscow, Russia}
\centerline{$^{24}$University of Nebraska, Lincoln, Nebraska 68588}
\centerline{$^{25}$New York University, New York, New York 10003}
\centerline{$^{26}$Northeastern University, Boston, Massachusetts 02115}
\centerline{$^{27}$Northern Illinois University, DeKalb, Illinois 60115}
\centerline{$^{28}$Northwestern University, Evanston, Illinois 60208}
\centerline{$^{29}$University of Notre Dame, Notre Dame, Indiana 46556}
\centerline{$^{30}$University of Panjab, Chandigarh 16-00-14, India}
\centerline{$^{31}$Institute for High Energy Physics, 142-284 Protvino, Russia}
\centerline{$^{32}$Purdue University, West Lafayette, Indiana 47907}
\centerline{$^{33}$Rice University, Houston, Texas 77251}
\centerline{$^{34}$University of Rochester, Rochester, New York 14627}
\centerline{$^{35}$CEA, DAPNIA/Service de Physique des Particules, CE-SACLAY,
                   France}
\centerline{$^{36}$State University of New York, Stony Brook, New York 11794}
\centerline{$^{37}$SSC Laboratory, Dallas, Texas 75237}
\centerline{$^{38}$Tata Institute of Fundamental Research,
                   Colaba, Bombay 400005, India}
\centerline{$^{39}$University of Texas, Arlington, Texas 76019}
\centerline{$^{40}$Texas A\&M University, College Station, Texas 77843}
}

\date{\today}

\maketitle

\begin{abstract}
We present new results on the search for the top quark in $p\bar p$
collisions at $\sqrt{s} = 1.8$ TeV with
an integrated luminosity of $13.5 \pm 1.6$ pb$^{-1}$.  We have considered
$t\bar t$ production in the Standard Model using electron and muon dilepton
decay channels ($t\bar t \rightarrow e \mu + \rm{jets}$,
$e e + \rm{jets}$, and $\mu \mu + \rm{jets}$) and
single-lepton decay channels ($t\bar t\rightarrow e + {\rm jets}$ and
$\mu + {\rm jets}$) with and without tagging of $b$ quark jets.
{}From all channels, we have 9 events with an expected background of
$3.8\pm0.9$.  If we assume that the excess is due to $t\bar t$ production,
and assuming a top mass of $180 \mbox{ GeV/c}^2$, we obtain a
cross section of $8.2 \pm 5.1$ pb.
\end{abstract}



 In the Standard Model (SM), the top quark is the weak isospin
 partner of the $b$ quark.  Precision electroweak measurements indirectly
 constrain the SM top quark mass to be
 $178 \pm 11^{+18}_{-19} \mbox{ GeV/c}^2$\cite{lepew}.
 The D\O\ collaboration recently published a lower limit on the mass of the top
 quark of $131 \mbox{ GeV/c}^2$, at a confidence level of
 95\%~\cite{dzerotop}.  The CDF collaboration has presented evidence
 for top quark production of mass
 $174 \pm 10^{+13}_{-12} \mbox{ GeV/c}^2$ with a
 cross section of $13.9^{+6.1}_{-4.8}$ pb~\cite{cdfnewtop}.
 The present
 analysis, which is based on the same data sample as
 Ref.~\cite{dzerotop}, includes three additional top quark decay channels,
 reoptimizes the event selection criteria for higher mass top, and
 provides a background-subtracted estimate of the
 top production cross section~\cite{luminosity}.

 We assume that the top quark is pair-produced and decays according to the
 minimal SM ({\it i.e.}\ $t\bar t \rightarrow W^+ W^-b\bar b$),
 and subsequently the $W$ decays into leptons
 ($W \rightarrow \ell \bar\nu$ where $\ell = e, \mu$ or $\tau$) or quarks.
 We searched for the
following distinct decay channels: (a)
$t\bar t \rightarrow \ell_1 \bar \ell_2 \bar \nu_1 \nu_2 b \bar b$,
the dilepton channels, with branching fractions 2/81 for
$e \mu + \rm{jets}$ and 1/81 each
for $e e + \rm{jets}$ and $\mu \mu + \rm{jets}$, and
(b) $t\bar t \rightarrow \ell \bar \nu q {\bar q}^\prime b \bar b$,
the single-lepton channels $e +\rm{jets}$ and
$\mu +\rm{jets}$,
each with a branching fraction of 12/81.
The latter were further subdivided into $b$-tagged and untagged
channels according to whether or not a soft muon was observed.
We denote the soft-$\mu$-tagged channels by
$e +\rm{jets}/\mu$ and $\mu +\rm{jets}/\mu$.
The probability that at least one of the two $b$
quarks in a $t\bar t$ event will decay to a muon, either directly or through a
cascade ($b\rightarrow c\rightarrow\mu$), is approximately 40\%~\cite{pdg}.


The D\O\ detector and data collection systems are described
in Ref.~\cite{dzeronim}.  The basic elements
of the trigger and reconstruction algorithms for jets, electrons,
muons, and neutrinos are
given in
Ref.~\cite{dzerotop}.

Muons were detected and momentum-analyzed using an iron toroid spectrometer
outside of a uranium-liquid argon calorimeter and a
non-magnetic central tracking system inside the calorimeter.
Muons were identified by their ability to penetrate the calorimeter and
the spectrometer magnet yoke.
Two distinct types of muons were defined.
``High-$p_T$'' muons, which are predominantly from gauge boson decay, were
required to
be isolated from jet axes by distance $\Delta{\cal R} > 0.5$
in $\eta$-$\phi$ space
($\eta$ = pseudorapidity = $\tanh^{-1}(\cos\theta)$; $\theta,\phi$ = polar,
azimuthal angle), and
to have transverse momentum $p_T > 12 \mbox{ GeV/c}$.
``Soft'' muons, which are primarily from $b$, $c$ or $\pi/K$
decay, were required to be within distance $\Delta{\cal R} < 0.5$ of any
jet axis
or, alternatively, to have a $p_T$ less than the minimum of a high-$p_T$
muon.  The minimum $p_T$ for soft muons was 4 GeV/c and the maximum $\eta$
for both kinds of muons was 1.7.

Electrons were identified by their longitudinal and transverse shower
profile in the calorimeter, and were required to have
a matching track in the central tracking chambers.
The background from photon conversions was reduced relative to that in
Ref.~\cite{dzerotop} by the imposition of an ionization
($dE/dx$) criterion on the chamber track.
Electrons were required to have $|\eta| < 2.5$ and transverse energy
$E_T > 15 \mbox{ GeV}$.

Jets were reconstructed using a cone
algorithm of radius ${\cal R} = 0.5$.

The presence of neutrinos in the final state was inferred from
missing transverse energy ($\rlap{\kern0.25em/}E_T$).  The
calorimeter-only $\rlap{\kern0.25em/}E_T$
($\rlap{\kern0.25em/}E_T^{\rm cal}$) was determined from
energy deposition in the calorimeter
for $|\eta| < 4.5$.  The total $\rlap{\kern0.25em/}E_T$ was
determined by correcting
$\rlap{\kern0.25em/}E_T^{\rm cal}$ for
the measured $p_T$ of detected muons.


The acceptance for $t\bar t$ events was calculated for several top
masses using the {\sc isajet} event
generator~\cite{isajet} and a detector simulation based on the {\sc geant}
program~\cite{geant}.

Physics backgrounds (those having the same final state particles as
the signal) were estimated by Monte Carlo simulation or from a
combination of Monte Carlo
and data.  The instrumental
background from jets misidentified as electrons was estimated
entirely from data using the measured jet misidentification probability
(typically $2 \times 10^{-4}$).
Other backgrounds for muons ({\it e.g.}\ hadronic punchthrough
and cosmic rays) were found to be negligible for the signatures
in question.


The signature for dilepton channels was defined as having two
high-$p_T$ isolated leptons, two jets and large
$\rlap{\kern0.25em/}E_T$.  The selection criteria
are summarized in Table~\ref{cuts}.

Physics backgrounds to the dilepton channels stem mainly from
$Z$ and continuum Drell-Yan production
($Z,\gamma^* \rightarrow ee, \mu \mu$ and $\tau\tau$), vector boson pairs
($WW$, $WZ$), and heavy flavor ($b\bar b$ and $c\bar c$) jet production.  The
$e \mu + \rm{jets}$
and $e e + \rm{jets}$ channels have additional backgrounds from jets
misidentified as electrons.
The background estimates
obtained with the new selection criteria are
more than a factor of two lower than those of Ref.~\cite{dzerotop}
whereas the acceptance for the $t\bar t$ signal in the high mass region
($m_t > 130 \mbox{ GeV/c}^2$) is similar.
Table~\ref{all} shows the acceptances and the expected number of $t\bar t$
events
for four values of the top mass, the total expected background, and the
number of observed events.
The $e\mu$ event that passes all selections was discussed in
Ref.~\cite{dzerotop}.


The signature for the single-lepton channels was defined as having one
high-$p_T$ lepton, large $\rlap{\kern0.25em/}E_T$, and a minimum of
three jets (with soft-$\mu$-tag) or four jets (without tag).  For the untagged
channels, additional background rejection was achieved through
event shape criteria based on the aplanarity
of the jets in the laboratory frame, $\cal A$~\cite{aplan}, and on the scalar
sum of the $E_T$'s of the jets, which we call $H_T$.
The criteria used for selecting the single-lepton channels are shown
in Table~\ref{cuts}.  Only the jets that passed the criteria of
Table~\ref{cuts} were used in calculating $\cal A$ and $H_T$.

The main backgrounds to the single-lepton $t\bar t$ channels were from
$W + {\rm jets}$, $Z + {\rm jets}$ (for $\mu +\rm{jets}/\mu$), and
multijet events where one jet was misidentified as an isolated lepton.
The latter do not normally
have large $\rlap{\kern0.25em/}E_T$.  We estimated the
multijet background directly from data,
based on the joint probability of multijet events having
large $\rlap{\kern0.25em/}E_T$ and a jet being misidentified as a lepton.

\begin{figure}[p]
\vspace*{1in}
\vbox{
\epsfysize=6.5in \epsfbox{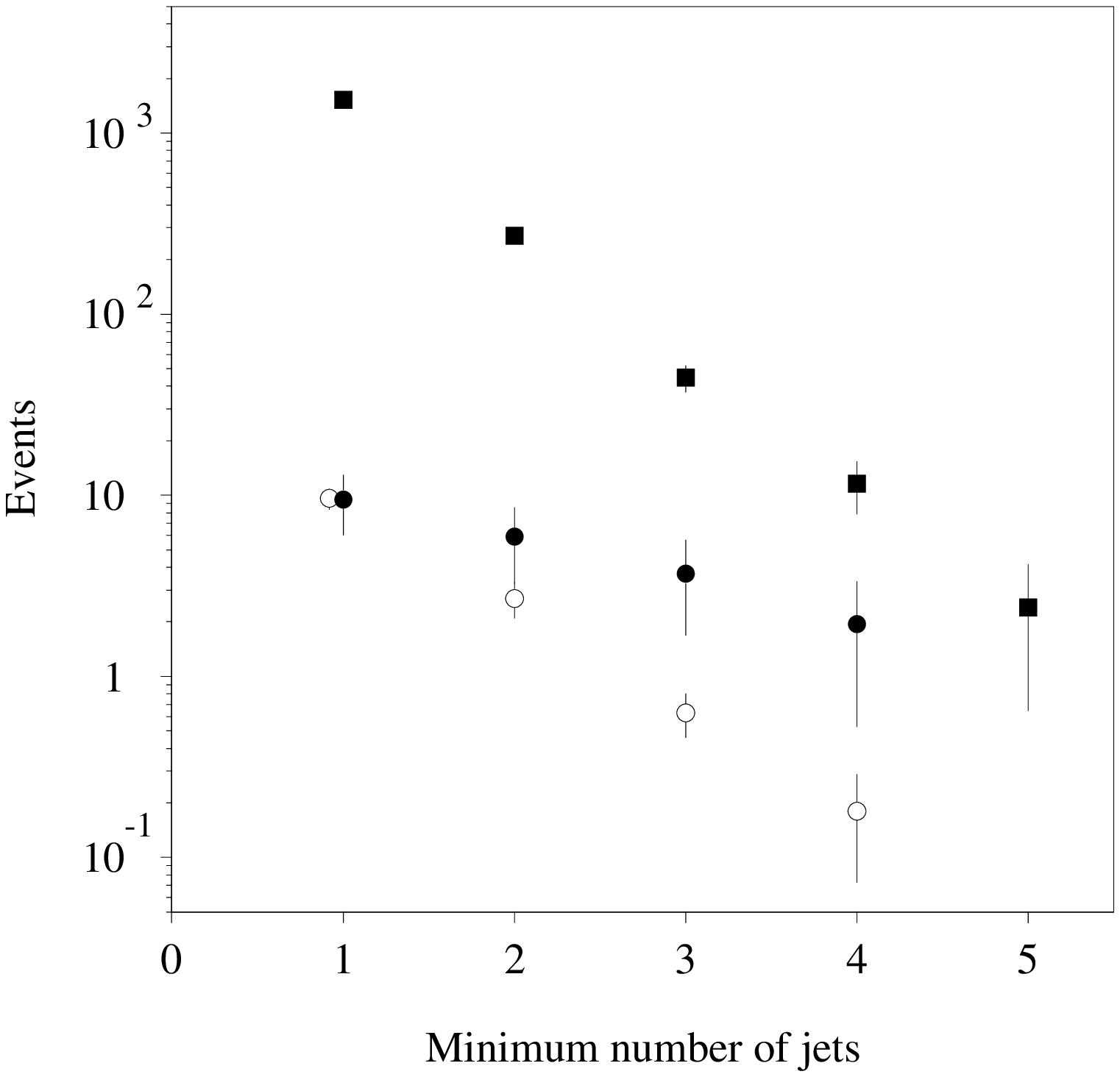}
\caption{The observed number of untagged (solid squares) and
soft-$\mu$-tagged (solid circles) $W + {\rm jets}$
events, after $Z$ and multijet background subtraction, and the number of
soft-$\mu$-tagged events expected in the absence of top (open circles),
as a function of the minimum number of jets.}
\label{wejets}
}
\end{figure}

Figure~\ref{wejets} shows the number of $Z$- and multijet-background-subtracted
$W + {\rm jets}$ events, for electrons and muons combined, as a function
of the minimum jet multiplicity, with and without soft-$\mu$-tag.
The $\rlap{\kern0.25em/}E_T$ and jet $E_T$ criteria were those of
Table~\ref{cuts} for the untagged and tagged analyses respectively.
Also shown is a prediction of the number of soft-$\mu$-tagged
$W + {\rm jets}$ events derived from the number of untagged
events folded with tagging rates obtained from multijet data.
The multijet tagging probability was observed to increase linearly
with
the number of jets, being approximately 1.5\% per
event for events with three or
more jets, and 2.0\% per event for events with four or more jets.
In contrast, the soft-$\mu$-tagging
probability for top quark events is calculated by Monte Carlo to be
about 20\% per event.
An admixture of a top signal in $W + {\rm jets}$ events could appear in high
jet multiplicities
(three or four jets) as excess untagged or tagged events.
In the absence of top, the number of $W + \rm jets$ events is expected to
decrease exponentially as a function of the jet multiplicity~\cite{berends}.
The observed number of tagged events is higher than the number
expected from background, but the significance of the excess is low.

\begin{figure}[p]
\vspace*{1in}
\vbox{
\epsfysize=6.5in \epsfbox{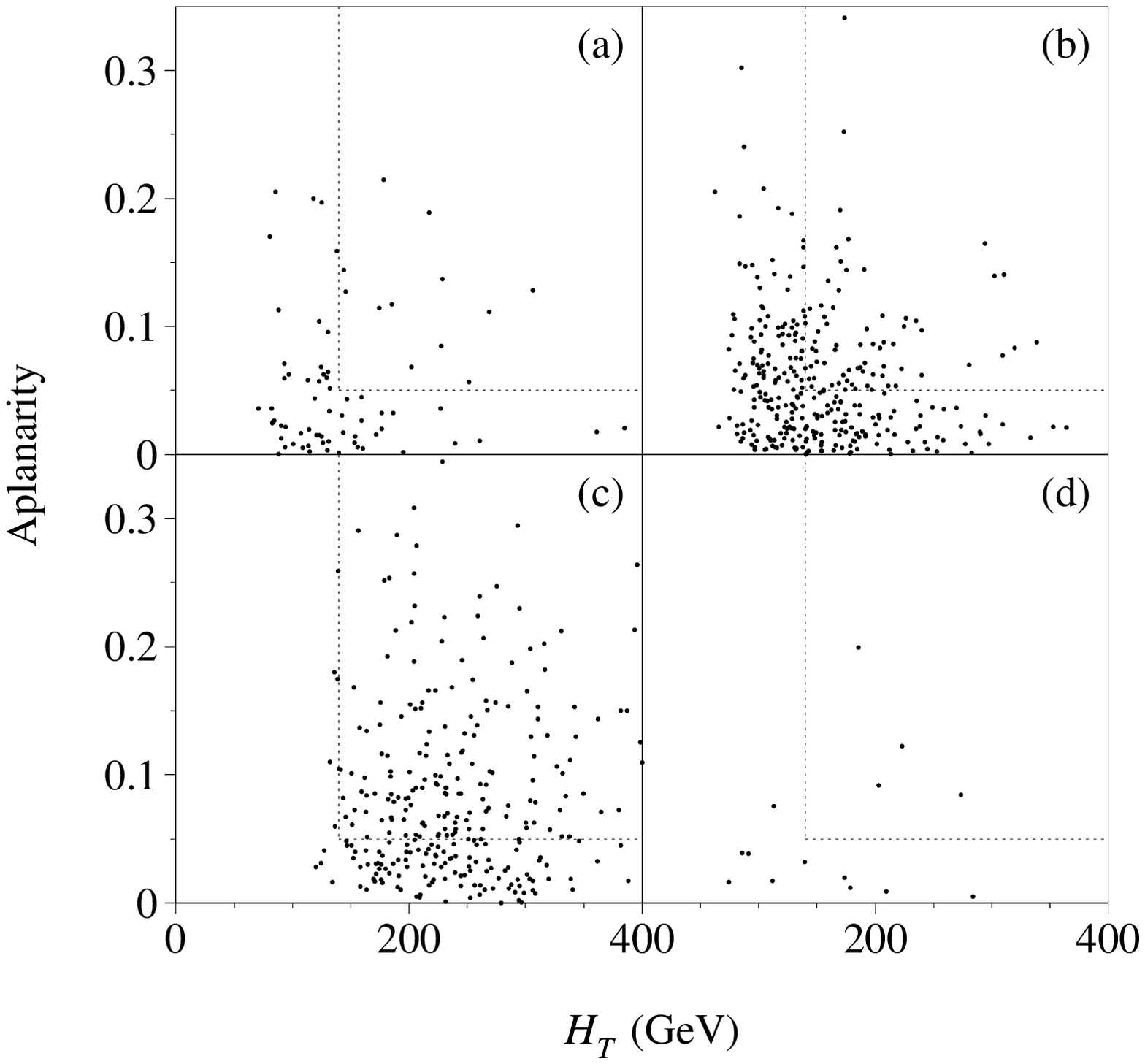}
\caption{$\cal A$ {\it vs.}~$H_T$ for single-lepton events for (a) multijet
background from data (effective luminosity = 60 $\times$ data luminosity),
(b) background from $W$ + 4 jet {\sc vecbos} Monte Carlo
(580 ${\rm pb}^{-1}$),
(c) 180 $\mbox{ GeV/c}^2$ top {\sc isajet} Monte Carlo
(2200 ${\rm pb}^{-1}$) and (d) data (13.5 ${\rm pb}^{-1}$).  The
dotted lines represent the event shape cuts used in the analysis.  }
\label{aht}
}
\end{figure}

The untagged single-lepton analysis made use of the distribution of
events in the $\cal A$-$H_T$ plane.  Figure~\ref{aht} shows
scatter plots of $\cal A$ {\it vs.}~$H_T$ for (a) expected multijet
background, (b) expected $W + \hbox{4 jet}$ background,
(c) $180 \mbox{ GeV/c}^2$ top, and (d) the observed
distribution for untagged
lepton + 4 jet events.  The multijet background was calculated from data
by the method described above.
The points in Fig.~\ref{aht}(a) are data points with a loosened
electron requirement, excluding real electrons.
The prediction for the untagged
$W + \hbox{4 jet}$
background (Fig.~\ref{aht}(b)) was calculated using the {\sc vecbos}
Monte Carlo
program~\cite{vecbos}.  The absolute normalization of the latter is not
well known due to theoretical uncertainties.  We normalized
the $W$ background directly from the data by two independent methods.
The first method was exponential extrapolation from one and two jets to four
jets.
The second method was to fit the observed distribution of events in the
entire $\cal A$-$H_T$ plane
(Fig.~\ref{aht}(d)) to a linear combination of signal and background
(Figs.~\ref{aht}(a)--(c)).  The
backgrounds and errors determined by the two methods agree
($1.9\pm0.7$ {\it vs.} $2.1 \pm 0.8$).
Our final background estimate is the average of the two methods.

Table~\ref{all} summarizes the results for all seven channels.
Adding all seven channels together, there are 9 observed events
with an expected background of $3.8\pm0.9$ events.  In the absence of top,
we calculate the probability of an upward fluctuation of the
background to 9 or more events to be 2.7\%.

\begin{figure}[p]
\vspace*{1in}
\vbox{
\epsfysize=6.5in \epsfbox{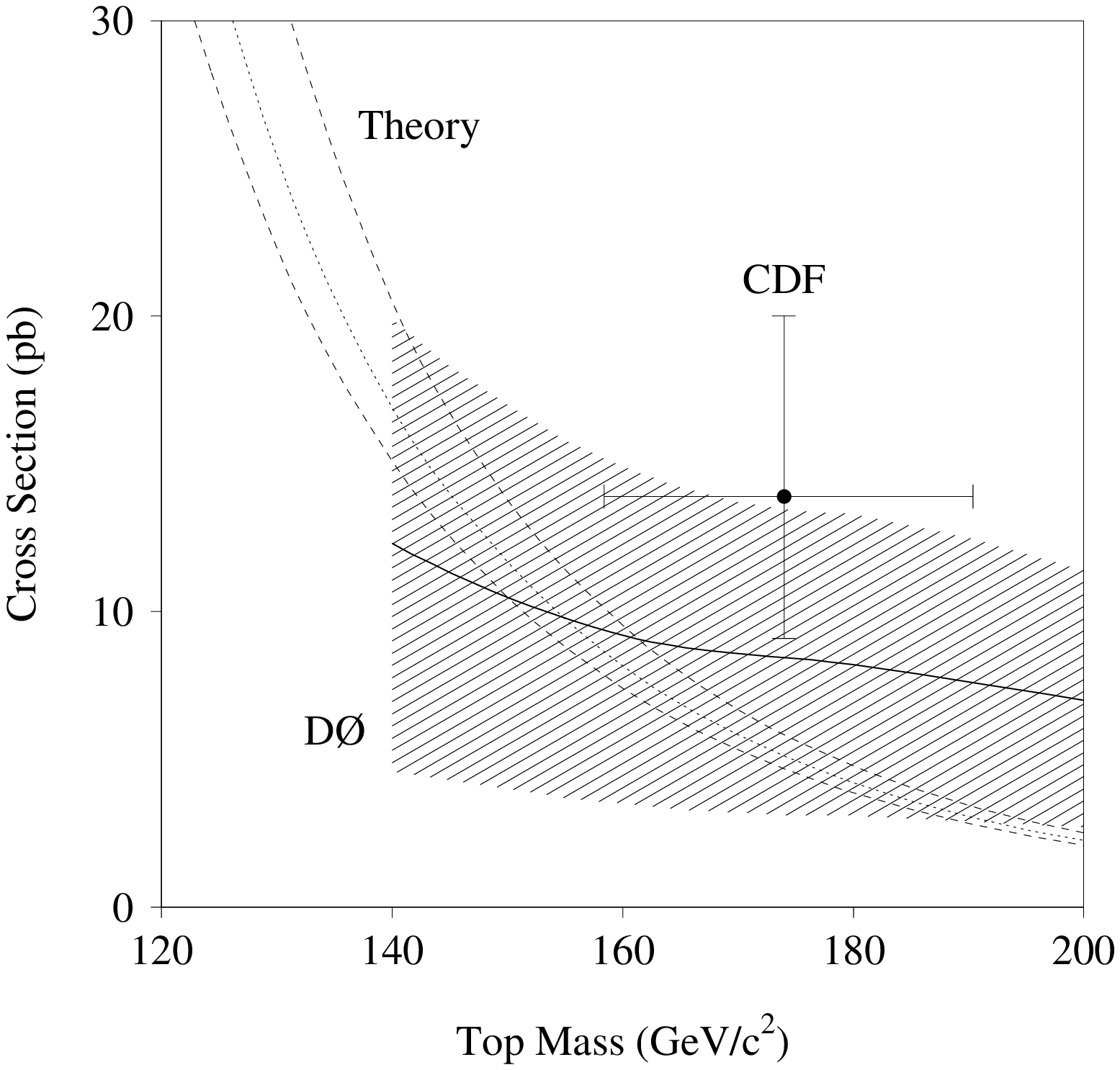}
\caption{D\O\ measured $t\bar t$ production cross section (solid line,
shaded band = one standard deviation error) as a function of top mass
 hypothesis.  Also
 shown are central (dotted line), high and low (dashed lines) theoretical
cross section curves~\protect\cite{laenen}, and the CDF
measurement~\protect\cite{cdfnewtop}.}
\label{limit}
}
\end{figure}

If we assume that the observed excess is due to $t\bar t$ production,
we can calculate the top cross section according to the equation
$
\sigma_{t\bar t} = {{\sum_{i=1}^7 { \left( N_i - B_i \right)}}
 / {\sum_{i=1}^7 {\varepsilon_i {\cal B}_i L_i}}}
$,
where $N_i$ is the number of observed events for decay channel $i$,
$B_i$ is the
expected background, $\varepsilon_i$ is the detection
efficiency for a particular top mass,
${\cal B}_i$ is the branching fraction, and $L_i$ is the
integrated luminosity.
The results are plotted in Figure~\ref{limit}.
For the 180 $\mbox{ GeV/c}^2$ (160~$\mbox{ GeV/c}^2$) top mass hypothesis, the
top production cross section is $8.2 \pm 5.1$ pb ($9.2 \pm 5.7$ pb).
This cross section is
consistent with theoretical expectations for the SM top
quark~\cite{laenen}.
Our measurement, although consistent with the CDF result~\cite{cdfnewtop}
and of comparable sensitivity, does not demonstrate the existence of the
top quark.


We thank the Fermilab Accelerator, Computing and Research Divisions, and
the support staffs at the collaborating institutions for their contributions
to the success of this work.   We also acknowledge the support of the
U.S. Department of Energy,
the U.S. National Science Foundation,
the Commissariat \`a L'Energie Atomique in France,
the Ministry for Atomic Energy in Russia,
CNPq in Brazil,
the Departments of Atomic Energy and Science and Education in India,
Colciencias in Colombia, CONACyT in Mexico,
and the Ministry of Education, Research Foundation and KOSEF in Korea.

\begin{table}[t]
\caption{Selection criteria for the seven top channels.}
\label{cuts}
\medskip
\begin{minipage}{4.25in}
\begin{tabular}{l|l}
$e \mu + \rm{jets}$ & $\ge 1$ electron ($E_T > 15 \mbox{ GeV}$,
$|\eta| < 2.5$) \\
& $\ge 1$ high-$p_T$ muon ($p_T > 12 \mbox{ GeV/c}$, $|\eta| < 1.7$) \\
& $\Delta {\cal R} (e,\mu) > 0.25$ \\
& $\ge 2$ jets ($E_T > 15 \mbox{ GeV}$, $|\eta| < 2.5$) \\
& $\rlap{\kern0.25em/}E_T^{\rm cal} > 20 \mbox{ GeV}$,
$\rlap{\kern0.25em/}E_T > 10 \mbox{ GeV}$ \\
\hline
$e e + \rm{jets}$ & $\ge 2$ electrons ($E_T > 20 \mbox{ GeV}$,
$|\eta| < 2.5$) \\
& $\ge 2$ jets ($E_T > 15 \mbox{ GeV}$, $|\eta| < 2.5$) \\
& $\rlap{\kern0.25em/}E_T^{\rm cal} > 25 \mbox{ GeV}$ \\
& $\rlap{\kern0.25em/}E_T^{\rm cal} > 40 \mbox{ GeV}$ if
$|m_{ee}-m_Z| < 12 \mbox{ GeV}$ \\
\hline
$\mu \mu + \rm{jets}$ & $\ge 2$ high-$p_T$ muons
($p_T > 15 \mbox{ GeV/c}$, $|\eta| < 1.1$) \\
& $\ge 2$ jets ($E_T > 15 \mbox{ GeV}$, $|\eta| < 2.5$) \\
& $m_{\mu \mu} > 10 \mbox{ GeV/c}^2$ \\
& $\Delta \phi^{\mu\mu} < 140^{\circ}$ if
$\rlap{\kern0.25em/}E_T < 40 \mbox{ GeV}$ \\
& $\Delta \phi (\rlap{\kern0.25em/}E_T^{\rm cal}, p_T^{\mu \mu} ) >
30^{\circ}$ \\
\hline
$e +\rm{jets}$ & 1 electron ($E_T > 20 \mbox{ GeV}$, $|\eta| < 2.0$) \\
& No soft muons \\
& $\ge 4$ jets ($E_T > 15 \mbox{ GeV}$, $|\eta| < 2.0$) \\
& $\rlap{\kern0.25em/}E_T^{\rm cal} > 25 \mbox{ GeV}$ \\
& ${\cal A} > 0.05$, $H_T > 140 \mbox{ GeV}$ \\
\hline
$\mu +\rm{jets}$ & 1 high-$p_T$ muon ($p_T > 15 \mbox{ GeV/c}$,
$|\eta| < 1.7$) \\
& No soft muons \\
& $\ge 4$ jets ($E_T > 15 \mbox{ GeV}$, $|\eta| < 2.0$) \\
& $\rlap{\kern0.25em/}E_T^{\rm cal} > 20 \mbox{ GeV}$,
$\rlap{\kern0.25em/}E_T > 20 \mbox{ GeV}$ \\
& ${\cal A} > 0.05$, $H_T > 140 \mbox{ GeV}$ \\
\hline
$e +\rm{jets}/\mu$ & 1 electron ($E_T > 20 \mbox{ GeV}$,
$|\eta| < 2.0$) \\
& $\ge 1$ soft muon ($p_T > 4 \mbox{ GeV/c}$, $|\eta| < 1.7$) \\
& $\ge 3$ jets ($E_T > 20 \mbox{ GeV}$, $|\eta| < 2.0$) \\
&  $\rlap{\kern0.25em/}E_T^{\rm cal} > 20 \mbox{ GeV}$ \\
&  $\rlap{\kern0.25em/}E_T^{\rm cal} > 35 \mbox{ GeV}$ if
$\Delta\phi(\rlap{\kern0.25em/}E_T^{\rm cal},\mu) < 25^\circ$ \\
\hline
$\mu +\rm{jets}/\mu$ & 1 high-$p_T$ muon ($p_T > 15 \mbox{ GeV/c}$,
$|\eta| < 1.7$) \\
& $\ge 1$ soft muon ($p_T > 4 \mbox{ GeV/c}$, $|\eta| < 1.7$) \\
& $\ge 3$ jets ($E_T > 20 \mbox{ GeV}$, $|\eta| < 2.0$) \\
& $\rlap{\kern0.25em/}E_T^{\rm cal} > 20 \mbox{ GeV}$,
$\rlap{\kern0.25em/}E_T > 20 \mbox{ GeV}$ \\
& For highest $p_T$ muon:
  $\Delta\phi(\rlap{\kern0.25em/}E_T ,\mu) < 170^\circ$ \\
& ~~~and
$\left| \Delta\phi(\rlap{\kern0.25em/}E_T , \mu) - 90^\circ
\right| / 90^\circ
< \rlap{\kern0.25em/}E_T / ( 45 \mbox{ GeV})$
\end{tabular}
\end{minipage}
\end{table}



\begin{table*}[t]
\caption{Efficiency $\times$ branching fraction
($\varepsilon \times {\cal B}$) and
the expected number of events ($\langle N \rangle$)
in the seven channels, based on the central theoretical $t\bar t$
production cross section of Ref.
{}~\protect\cite{laenen}, for four top masses.
Also given are the
expected backgrounds, integrated luminosity, and the number of
observed events in each channel.}
\label{all}
\medskip
\begin{tabular}[p]{cc|c|c|c|c|c|c|c|c}
\multicolumn{2}{c|}{$m_t$ (GeV/$c^2$)} &
\multicolumn{1}{c|}{$e \mu + \rm{jets}$} &
\multicolumn{1}{c|}{$e e + \rm{jets}$~~} &
\multicolumn{1}{c|}{$\mu \mu + \rm{jets}$~~~} &
\multicolumn{1}{c|}{$e + \rm jets$~} &
\multicolumn{1}{c|}{$\mu+\rm jets$~ } &
\multicolumn{1}{c|}{$e + \rm jets /\mu$} &
\multicolumn{1}{c|}{$\mu+\rm jets /\mu$} &
\multicolumn{1}{c}{ALL} \\
\hline
\hline
      & $\varepsilon \times {\cal B} (\%) $ & $0.31 \pm 0.04$ & $0.18\pm 0.02$
      & $0.15\pm 0.02$ & $1.1\pm 0.3$ & $0.8\pm 0.2$ &  $0.6\pm 0.2$ &
        $0.4\pm 0.1$ & \\
            \cline{2-10}
 ~140 & $\langle N \rangle$             & $0.71 \pm 0.12$ & $0.41\pm 0.07$
      & $0.25\pm 0.04$ & $2.5\pm 0.7$ & $1.3 \pm 0.4$  &   $1.4 \pm 0.5$
      & $0.7\pm 0.2$ & $7.2\pm 1.3$ \\
\hline
      & $\varepsilon \times {\cal B} (\%) $ & $0.36 \pm 0.05$ & $0.20\pm 0.03$
      & $0.15\pm 0.02$ & $1.5\pm 0.4$ & $1.1\pm 0.3$ &  $0.9\pm 0.2$ &
        $0.5\pm 0.1$ & \\
            \cline{2-10}
 ~160 & $\langle N \rangle$             & $0.40 \pm 0.07$ & $0.22\pm 0.04$
      & $0.12\pm 0.02$ & $1.7\pm 0.5$ & $0.9 \pm 0.3$  &   $1.0 \pm 0.3$
      & $0.4\pm 0.1$ & $4.7\pm 0.8$ \\
\hline
      & $\varepsilon \times {\cal B} (\%) $ & $0.39 \pm 0.05$ & $0.21\pm 0.03$
      & $0.14\pm 0.02$ & $1.6\pm 0.4$ & $1.1\pm 0.3$ &  $1.1\pm 0.2$ &
        $0.7\pm 0.1$ & \\
            \cline{2-10}
 ~180 & $\langle N \rangle$             & $0.22 \pm 0.04$ & $0.12\pm 0.02$
      & $0.06\pm 0.01$ & $0.9\pm 0.3$ & $0.5 \pm 0.1$  &   $0.6 \pm 0.1$ &
        $0.3\pm 0.1$ & $2.7\pm 0.4$ \\
\hline
      & $\varepsilon \times {\cal B} (\%) $ & $0.40 \pm 0.05$ & $0.30\pm 0.04$
      & $0.14\pm 0.02$ & $1.8\pm 0.4$ & $1.3\pm 0.3$ &  $1.4\pm 0.1$ &
        $0.8\pm 0.2$ & \\
            \cline{2-10}
 ~200 & $\langle N \rangle$             & $0.12 \pm 0.02$ & $0.09\pm 0.02$
      & $0.03\pm 0.01$ & $0.5\pm 0.1$ & $0.3 \pm 0.1$  &   $0.4 \pm 0.1$ &
        $0.2\pm 0.1$ & $1.7\pm 0.3$ \\
\hline
\hline
\multicolumn{2}{c|}{Background}    & $0.27 \pm 0.06$ & $0.16 \pm 0.07$ &
$0.33 \pm 0.06$ & $1.3 \pm 0.7$ & $0.7 \pm 0.5$ & $0.6 \pm 0.2$  &
$0.4 \pm 0.1$ & $3.8 \pm 0.9$ \\
\hline
\multicolumn{2}{c|}{$\int {\cal L}dt \ (\rm pb^{-1})$} & $13.5 \pm 1.6$ &
$13.5 \pm 1.6$
& $9.8 \pm 1.2$ & $13.5 \pm 1.6$ & $9.8 \pm 1.2$ & $13.5 \pm 1.6$ &
$9.8 \pm 1.2$ & \\
\hline
\multicolumn{2}{c|}{Data} & \multicolumn{1}{c|}{1} &
\multicolumn{1}{c|}{0} & \multicolumn{1}{c|}{0} &
\multicolumn{1}{c|}{2} & \multicolumn{1}{c|}{2} &
\multicolumn{1}{c|}{2} &  \multicolumn{1}{c|}{2} &
\multicolumn{1}{c}{9}
\end{tabular}
\end{table*}



\end{document}